# How libraries classified physics preprints before arXiv and set the stage for distinguishing insiders from outsiders

Phillip H. Roth
RWTH Aachen University
phillip.roth@khk.rwth-aachen.de

*In a world with ever-growing scientific literature, meaningful classifications are vital to keep on top of the latest results. In this Comment, historian and sociologist Phillip Roth traces the history of preprint classification in physics.*

"The correct categorization of scholarly works is vitally important" stated the arXiv blog[i], as it announced a new user-driven classification tool in 2020. As the post continues to explain, classification of submissions into subfields of physics, like condensed matter (*cond-mat*) or high energy physics phenomenology (*hep-ph*), "helps readers find the information they seek – and assures authors that their research will be found by the right audience." Categories and classification systems help arrange the world, including the world of science and its literature, in a meaningful way. But what defines correct categorization?

Ideally, correct categories would derive from a theoretical system that clusters the available literature in the field around scientifically meaningful concepts. However, in practice, classification has often relied on ad-hoc approaches, combined with some organized deliberations and international coordination.[ii] Looking back at the history of classification of preprints in physics reveals policies that revolved around the informational needs of physics researchers and pragmatic solutions to share knowledge around the world, as quickly as possible.

Scholars in the social sciences and information science have emphasized how classification systems hide behind a veneer of technicality the political and social struggles that went into their formation and application.[iii] While categories appear as neutral labels, their relationship to each other, as well as to the items they denote, reveal underlying tensions between an ideal systematization and technical solutions for managing the flood of scientific information. The handling of preprints as we know it today was the invention of a few leading physics libraries, who processed them primarily to the benefit of their local communities. As these methods and categorize became institutionalized, they helped create the strong boundary that seperates insiders from outsiders in the field.

## From private circulation to public display

At first, preprints were only circulated between private networks of physicists. In the post-World War II era, journal publication was picking up as a mode of sharing results, but there were often long publication times associated with getting an article into print, so physicists would directly share notes and manuscripts with colleagues around the globe, to stay abreast of the newest developments in the field[iv]. Belonging to the class of so-called "grey literature," librarians as early advocates of formal preprint handling called them "temporary documents" to share vital information with the right audience as soon as possible, while papers underwent the long process of formal publication in a journal.[v]

It was in the field of high-energy physics that preprints began to be collected and categorized by libraries. The reason was practical. An innovative librarian at CERN, Luisella Goldschmidt-Clermont, had observed that many physicists who came to the Geneva laboratory for a short visit often became out of touch with their usual informational networks. Preprints may be sent to their home institutions rather than to CERN, making it difficult for them to stay on top of the latest results. Her solution was to gather the preprints at CERN directly.

In the late 1950s, the library at CERN began soliciting preprints from physicists, asking them to send their papers to the library, instead of to their private mailing networks. The librarians would then catalogue the preprints and put them on public display in the reading room. By centrally documenting the preprints, making this formerly-private information now public, the CERN library created the first formal preprint exchange system, thereby paving the way for many defining practices of preprint handling as they live on today on arXiv and other preprint servers.[vi]

CERN was not the only library to implement systems for documenting preprints in circulation. In 1962, the CERN method of preprint handling came to the library at the Stanford Linear Accelerator (SLAC); and in 1969, initiated by another innovative librarian, Louise Addis, SLAC began issuing a weekly newsletter "Preprints in Particles and Fields," an accession list which brought information about the newest preprints in high-energy physics to subscribers around the globe.

Meanwhile, the *High-Energy Physics Index* (HEPI) was published by the German Electron Synchrotron (DESY) in Hamburg beginning in 1963. The HEPI was a biweekly publication containing a bibliography of recent preprints, conference proceedings, and conventional literature. In its heydays, before the introduction of desktop computer publishing and networked communication in the 1980s, the HEPI had more than 300 (mostly institutional) subscribers worldwide. As one of the fastest and most complete forms of documentation of the physics literature, it was a valuable information source in high-energy physics before the advent of arXiv or the INSPIRE online database.

**A need for selectivity and categorization**

Soon, the growing influx of preprints, as well as the increasing information demands of researchers, required that library staff improve their handling and sorting of the newest accssions. At CERN, where the demand for preprints increased by almost 300% between 1964 and 1967,[vii] the library began to understand itself as an important gatekeeper of the "grey" physics liteature. "The usefulness of a special library," the 1965 Staff Manual states,

"depends in large measure on its selectivity," while it warns that a "heterogeneous mass of vaguely related documentation can choke or crowed out the relevant and important items."[viii] Libraries were often aided by physicists in making their selections, or employed "scientific information officers," people trained in both physics and librarianship, who scanned papers sent in and helped categorize them – much in the way that content moderation is today performed on preprint servers.

Next to formal requirements about langauge and style, the CERN library introduced a pragmatic categorization system to make the stream of preprints manageable. Papers were given simple subject categories from the field of high-energy physics, such as "theoretical particle physics," "high energy experimental physics," "experimental techniques," "detectors," or "accelerators," purely to enable the list to be sorted in a hierarchical order. As such lists gained in popularity, scientists not only used them as an information resource, but increasingly also strove to be included in libraries' preprints lists themselves. Inclusion in a preprint list came to act as a badge of membership in the field and offered an opportunity to expose one's work to the research community. This sort of social incentive continues today with preprint servers, where authors have been found to "race to submit their most important papers at the right moment in order to meet the need for increased visibility."[ix]

### Not all categories are created equal

The categories of the CERN library's list did not represent neutral classes, though. Instead, they encoded in the subject classification mainly topics of interest for the research conducted at the Geneva laboratory. Success of one's exposure thus hinged on conforming to the research interests and trendiness of subjects as they were seen by CERN scientists. DESY found another route to ordering the growing stack of preprints. The HEPI was designed with an extensive subject keyword, author, and report series index to make submissions retrievable for users – essentially, the first abstracting service for the grey literature. Subject keywords were assigned by physicists based on a keyword list regularly updated at DESY to map the field of research in high-energy physics.[x] However, this form of classification also did not ensure neutrality. The more mainstream a contribution was, the more keywords were assigned; and the more keywords a paper had assigned, the easier it was to be found and recognized by fellow physicists. At the same time, the fewer keywords a paper had, the less relevant it might have appeared in the eyes of the physics audience.

On arXiv, the assignment of unequal relevance is hidden behind the general physics (*gen-ph*) class. The website uses a set of categories that organize preprint submissions into specialized classes such as astrophysics, condensed matter, or specific categories for high-energy physics, as subject calssification becomes vitally important for search and retrieval in the digital age.[xi] Although the general physics category appears as just another specialist class next to others, "*gen-ph* was made to contain papers that have no specific audience and are therefore thought to be generally uninteresting to all specialized audiences."[xii] As the website reserves itself the right to reclassify submissions into the *gen-ph* category, it performs a form of social sorting that deems certain papers less relevant for the professional communities represented on arXiv.

The library at CERN recognized early on that such forms of content moderation introduced "certain dangers." As the manual admits, "the rejection of 'border' material is inevitably somewhat arbitrary." Today, this arbitrariness is extended from humans to machines. While arXiv's user-driven and machine learning-based classification tool is meant to lift the burden off moderators and to "empower authors and increase transparency," everytime the system recommends to reclassify a submission, it reveals a boundary not drawn according to an ideal representation of physics, but as the outcome of practices and methods that have become naturalized over time and enforce human visions of separating insiders from outsiders.

**Competing interests**

The author declares no competing interests

---